\documentclass[pdflatex,sn-basic]{sn-jnl}

\usepackage{graphicx}%
\usepackage{multirow}%
\usepackage{amsmath,amssymb,amsfonts}%
\usepackage{amsthm}%
\usepackage{mathrsfs}%
\usepackage[title]{appendix}%
\usepackage{xcolor}%
\usepackage{textcomp}%
\usepackage{manyfoot}%
\usepackage{booktabs}%
\usepackage{algorithm}%
\usepackage{algorithmicx}%
\usepackage{algpseudocode}%
\usepackage{listings}%
\usepackage[utf8]{inputenc}
\DeclareMathOperator*{\argmax}{arg\,max}

\theoremstyle{thmstyleone}%
\theoremstyle{thmstyletwo}%

\theoremstyle{thmstylethree}%

\begin{document}

\title[Deploying Robust Decision Support Systems for Transit Headway Control: Rider Impacts, Human Factors and Recommendations for Scalability]{Deploying Robust Decision Support Systems for Transit Headway Control: Rider Impacts, Human Factors and Recommendations for Scalability}


\author*[1]{\fnm{Joseph} \sur{Rodriguez}}\email{rodriguez.josep@northeastern.edu}

\author[1]{\fnm{Haris N.} \sur{Koutsopoulos}}\email{h.koutsopoulos@northeastern.edu}

\author[2]{\fnm{Jinhua} \sur{Zhao}}\email{jinhua@mit.edu}

\affil*[1]{\orgdiv{Department of Civil and Environmental Engineering}, \orgname{Northeastern University}, \orgaddress{\street{360 Huntington Ave}, \city{Boston}, \postcode{02115}, \state{MA}, \country{USA}}}

\affil[2]{\orgdiv{Department of Urban Studies and Planning}, \orgname{Massachusetts Institute of Technology}, \orgaddress{\street{77 Massachusetts Ave}, \city{Cambridge}, \postcode{02139}, \state{MA}, \country{USA}}}








\abstract{Service reliability is critical to transit service delivery. This paper describes headway control pilots conducted in two high-ridership Chicago bus routes between 2022 and 2023. A decision support system was developed for a bus holding strategy based on a reinforcement learning approach. For the pilots, a user interface enabled supervisors to monitor service and record applied actions. The first pilot tested terminal-based holding on a route affected by missed trips from absenteeism. The analysis found improvements in reliability, and the application of control was shown to outperform days with more service. The second pilot applied en-route holding in a high-ridership bus route in Chicago. The evaluation showed wait time improvements with rippled benefits to stops downstream, and a reduction in transfer times from connecting bus and rail lines. Compliance analysis based on the supervisor logs on the app revealed mixed compliance levels from drivers, which were related to the mentality of schedule adherence and seniority. Recommendations are provided for practitioners to scale similar efforts.}



\maketitle

\section{Introduction}\label{intro}
Service reliability is critical to transit service delivery. Riders of high-frequency services typically arrive at a stop without checking the schedule and expect the bus to arrive. Even for operations with a well-optimized schedule, mixed-traffic operations, uncertain rider demand and other challenges on the field force plans to change. Thus, operations control serves as a complementary practice to deal with deviations from the plan.\par 

Typically, operations staff at agencies are tasked with applying proactive and corrective control actions to maintain service regularity. Historically, these tasks are divided among central controllers and field supervisors stationed at terminals or mid-route points. However, these decision-makers can have varying levels of expertise and access to information. Human decision-making in the field is shaped by these factors—meaning that two supervisors in similar circumstances can produce very different outcomes. Data-driven decision support systems can help make the information available and the criteria for decisions more consistent across operational actors.\par

Real-time operations control has been the subject of much theoretical and experimental work. At the theoretical level, methods for control strategies range from heuristics to advanced optimization and machine learning methods \cite{tirachini_headway_2021}. Despite the abundance of theoretical models for real-time control and advanced technologies, these have not been broadly adopted in real operations at transit agencies. Control decisions at transit agencies remain largely ad-hoc, manual, and under-digitized. Only the transportation authorities of  Stockholm, Sweden  \cite{cats_regularity-driven_2014}, and Santiago, Chile \cite{lizana_bus_2014} have reported reliance on real-time tools for headway-based control. In real-world pilots, the methods have been limited to heuristics and rolling-horizon optimization, with the computational load limiting the practicality for deployments. These methods, which rely on assumptions and rules as opposed to data-driven training, are limited in their ability to generalize to real-world conditions and constraints, such as driver non-compliance. Human factors play a significant role in the effectiveness of holding strategies, as evidenced in simulation studies \cite{phillips_quantifying_2015} and driver surveys \cite{martinez-estupinan_understanding_2022, chen_bus_2024}. Such issues are seldom accounted for in the development and evaluation of proposed methods, but can impact the effectiveness of deployments. However, the literature has seldom considered driver behavior explicitly in the deployed algorithms. \cite{argote-cabanero_dynamic_2015} constrained the variability of holding instructions displayed to drivers to avoid driver stress and improve compliance, thus embedding human factors into the control design. The method, however, does not fully capture the intrinsic motivations of drivers not to comply, such as the culture of adhering to schedules. \par

To address this, we deploy a data-driven reinforcement learning (RL) method for terminal and en-route holding in two pilots, with human factors embedded in the methodology. The models are pre-trained in simulated operational conditions that are based on empirical data sources. For robust deployments, guardrails are built around the model's recommended holding actions to enforce service plan constraints, such as limits on lateness and excessive holding times. Also, an even-headway strategy similar to classical methods is formulated and used as an additional source of holding recommendations, as an alternative in case the RL model fails or produces suboptimal recommendations. This recommendations engine is integrated into a decision support system for transit supervisors and controllers. The decision support system bridges existing information gaps between stakeholders of real-time service decisions. These stakeholders, such as field supervisors, the control center and garage staff, typically have different information access and needs without an integrated information infrastructure, resulting in an over-reliance on verbal communication. The decision support system combines automated information pipelines from various sources with manual service update features to allow the latest information to be visible to stakeholders. The human factors are embedded in the methodology in various ways. First, the RL model training includes simulated scenarios with driver non-compliance. Second, the decision support system allows supervisors to report holding instructions, which allow the tracking of non-compliance. Lastly, the analysis advances the understanding of the relationship between non-compliance, schedule adherence, and experience levels. \par

To summarize, the main contributions of this paper are:
\begin{itemize}
    \item To the authors' best knowledge, this study represents the first application of a reinforcement learning model in a live deployment of bus operations control.
    \item The paper embeds human factors in the methodology, with the models trained on non-compliant drivers and tracking of compliance. It also quantifies the compliance levels and explores the relationship with driver attributes.
    \item This paper presents a comprehensive impact assessment with metrics relevant to riders, bus drivers, and transit agency stakeholders. It includes comprehensive performance metrics, including impact on transfer times at key interchanges and effectiveness relative to adding more service.
\end{itemize}

The remainder of this paper is structured as follows. First, an overview of the related work is provided, followed by a description of the methodology and the decision support system. Afterwards, the implementation and performance impacts from two pilots are discussed. Next, recommendations for scalability are provided, as well as conclusions.

\section{Related Work}\label{sec5}

\subsection{Previous deployments}
Academic collaborations with transit agencies have yielded numerous pilots and long-term implementations of headway control. The papers on previous deployments are synthesized along a set of dimensions in Table \ref{tab:lit-review}. The deployments go back decades and have mostly been based in major US cities, with some in Europe and one in Chile. Most applications are on bus systems, with few on rail, potentially due to the flexibility and lower ridership volume of bus services. Pilots typically apply a single strategy, most commonly holding, with one study applying short-turning \cite{strathman_bus_2001} and several studies applying speed controls \cite{cats_regularity-driven_2014, lizana_bus_2014, argote-cabanero_dynamic_2015}. In several implementations, the control actions were decided by a manual process, without reliance on an automated decision support system (DSS). However, recent work has produced advanced DSS with user interfaces. The dominant form of communication with the driver happens in person and by radio. In contrast, experiments in Stockholm \cite{cats_regularity-driven_2014}, Santiago \cite{lizana_bus_2014}, Atlanta \cite{berrebi_translating_2018}, and San Sebastian, \cite{argote-cabanero_dynamic_2015} used on-board guidance for drivers to regulate headways.\par

Most studies neglect the impact assessment on cycle times, which is a critical performance indicator at agencies. The impact on transfer times has also not been addressed. Transfer times are representative of the rider experience and can be measured using current data sources, such as Automatic Vehicle Location (AVL) and Automated Fare Collection (AFC) data. Additionally, there is limited data on the levels of driver compliance with control strategies. Finally, control methods used in the literature are limited to heuristics and predictive control approaches, which may not adapt well to certain real-world conditions such as driver non-compliance. No techniques that train with empirical data, such as machine learning methods, which may capture real-world conditions more accurately, have been applied in practice.\par 

\begin{sidewaystable}
\caption{Summary of previous deployments of transit headway control}\label{tab:lit-review}
\tiny
\begin{tabular*}{\textheight}{p{1.2cm}p{1.2cm}p{1cm}p{1cm}p{1cm}ccccccccc}
\toprule
 & & & \multicolumn{4}{@{}c@{}}{Control Strategy} & & \multicolumn{5}{@{}c@{}}{Evaluation factors}\\
 \cmidrule{5-7}\cmidrule{9-14}
Study & Location & Duration & Mode & Holding & Short-turn & Speed & DSS & Regularity & Crowding & Wait time & Cycle time & Compliance & Transfer \\
\midrule
\cite{englisher_minneapolis-st_1984} & Minneapolis - St. Paul & 2 weeks & Bus & \checkmark & & & & \checkmark & & \checkmark & & \\
\cite{abkowitz_implementing_1990} & Boston & 1 week & Bus & \checkmark & & & & \checkmark & \checkmark & & \checkmark & \\
\cite{strathman_bus_2001} & Portland & 3 weeks & Bus & \checkmark & \checkmark & & & \checkmark & \checkmark & & & \\
\cite{pangilinan_bus_2008} & Chicago & <1 week & Bus & \checkmark & & & & \checkmark  & & & & & \\ 
\cite{bartholdi_iii_self-coordinating_2012} & Atlanta & <1 week & Bus & \checkmark & & & & \checkmark & & & & & \\
\cite{cats_regularity-driven_2014} & Stockholm & Long-term & Bus & \checkmark & & \checkmark & & \checkmark & \checkmark & & & & \\
\cite{lizana_bus_2014} & Santiago & <1 week & Bus & \checkmark & & \checkmark & \checkmark & \checkmark & & & & & \\
\cite{argote-cabanero_dynamic_2015} & San Sebastian & Long-term & Bus & \checkmark & & \checkmark & \checkmark & \checkmark & & & & \checkmark & \\
\cite{maltzan_using_2015} & Boston & 1 week & Bus & \checkmark & & & \checkmark & \checkmark & & \checkmark & & \checkmark & \\
\cite{fabian_improving_2018} & Boston & 1 week & Rail & \checkmark & & & \checkmark & \checkmark & & \checkmark & \checkmark & \checkmark & \\
\cite{berrebi_translating_2018} & Atlanta & 2 weeks & Bus, Rail & \checkmark & & & \checkmark & \checkmark & & \checkmark & & & \\
\cite{wood_real-time_2018} & New York & Long-term & Rail & \checkmark & & & \checkmark & & & & & & \\
\cite{soza-parra_lessons_2019} & Washington, D.C. & 4 weeks & Bus & \checkmark & \checkmark & & & \checkmark & & \checkmark & & & \\
\cite{wolofsky_towards_2019} & Boston & 2 weeks & Rail & \checkmark & & & & \checkmark & \checkmark & \checkmark & \checkmark & & \\
Rodriguez et. al (2025) & Chicago & 3 weeks & Bus & \checkmark & \checkmark & & \checkmark & \checkmark & \checkmark & \checkmark & \checkmark & \checkmark & \checkmark\\
\bottomrule
\end{tabular*}
\end{sidewaystable}

\subsection{Human factors}
Recent work has explored how varying driver behaviors impact the effectiveness of control strategies. \cite{maltzan_using_2015} found that a minority of bus drivers were responsible for a large portion of early and late terminal departures, using Boston MBTA drivers as a case study. In more recent work, \cite{martinez-estupinan_improving_2023} found significant heterogeneity in the driving speed of bus drivers in Chile and showed its impacts on the effectiveness of headway control.\par 

The phenomenon of driver non-compliance with control instructions has also been studied. \cite{martinez-estupinan_understanding_2022} conducted a survey of drivers in Chile regarding their perception of on-board headway control instructions. A critical finding was that experienced drivers reported lower compliance with instructions. The relationship between seniority and compliance has not yet been studied empirically. \cite{chen_bus_2024} also conducted a survey of drivers to ask about their perceptions of the benefits of control and their compliance intentions. They found that higher mental workload and confidence were associated with lower intentions to comply. The authors also found that gender was associated with compliance, with female drivers being more compliant. In terms of empirical analysis, \cite{fabian_improving_2017} measured compliance during a pilot implementation of terminal-based holding on a rail line in Boston. The authors found that operators tend to comply less when the recommendations are further from the scheduled time. To assess the systematic impacts of non-compliance, \cite{phillips_quantifying_2015} conducted a simulation study of holding strategies under various scenarios of non-compliance. In the first set of scenarios, every vehicle had the same probability of compliance with holding instructions. The results showed that in this case, the sensitivity of the performance to non-compliance was low, with 88\% of the maximum potential benefits achieved even with 50\% compliance. The second set of scenarios tested permanently non-compliant vehicles, which resulted in a substantial drop in performance. These findings highlight the potential of overestimated improvements in theoretical studies that neglect driver behavior.
 
\section{Control Methods}\label{sec6}
This section formulates the holding problem for a single route, generalized to include holding decisions at terminal or mid-route stops. Subsequently, it describes the approach using even-headway and reinforcement learning strategies. \par

\subsection{Holding problem}

Consider the arrival of trip $i$ at stop $j \text{ where } j \in \{0, 1, 2, \ldots, M-1\}$ and $0$ and $M$ are the start and end terminals, respectively. The departure time at a stop is determined by the dwelling time, passenger activity, schedule-related constraints, and the holding time:
\begin{equation}
    DT_{ij} = DT_{ij}^{min} + HT_{ij}
\end{equation}
Where $DT_{ij}$ is the departure time, $DT_{ij}^{min}$ is the earliest departure time and $HT_{ij}$ is the holding time of trip $i$ at stop $j$. It should be noted that holding is typically applied to a set of pre-defined control stops. $DT^{min}_{ij}$ is defined separately for the terminal and mid-route stops, as follows: 
\begin{equation}
    DT_{ij}^{min} = \begin{cases}
\max(ST_{ij} - s^{e}, AT_{ij} + PT_{ij} + L^{\min}) & \text{if } j = 0 \\
 AT_{ij} + PT_{ij} & \text{if } j > 0
\end{cases}
\end{equation}
Where $ST_{ij}$ is the scheduled departure time for trip $i$ at stop $j$, $s^e$ is the schedule deviation threshold for early departures, $AT_{ij}$ is the arrival time, $PT_{ij}$ is the passenger-related dwell time and $L^{min}$ is the minimum layover time. At the terminal, $ST_{ij} - s^{e}$ is enforced as the earliest possible departure time, where $s^e$ is set by the standards of the respective agency. $s^e=0$ would prevent early departures altogether and could be beneficial to passengers who rely on the schedule to not miss their bus. Conversely, for $s^e>0$ , holding times could result in early departures. This schedule relaxation can be considered by agencies, given that riders of high-frequency services tend to rely less on schedules. Furthermore, early departures may be preferred by drivers compared to late departures that may be perceived as disruptive to their work schedule.\par

The minimum layover time $L^{min}$ is another policy-set standard for drivers to have a guaranteed minimum layover time. If sufficient schedule padding is added to the scheduled cycle time, most trips have longer than $L^{min}$ for terminal layover time, and thus it does not translate into a delay for the next trip. $PT_{ij}$ encapsulates the time for the vehicle to stop, board and/or alight passengers, and, if $j>0$, get back on the route.\par

Given the conditions of the problem, the holding time is determined by the control strategy and is constrained by the maximum schedule lateness by the agency's policies, formulated as follows:

\begin{equation}
    HT_{ij} = \begin{cases}
0 & \text{if } DT_{ij}^{min} > ST_{ij} + s^{l} \\
 \pi(\cdot) & \text{otherwise} 
\end{cases}
\end{equation}

Where $s^l$ is the maximum late schedule deviation, and  $\pi(\cdot)$ is the holding time policy. Similar to $s^e$, $s^l$ is determined according to the agency standards on lateness. $\pi(\cdot)$ is defined according to the selected methodology and corresponding input variables. The remainder of this section details the two methodologies considered in the pilots.

\subsection{Even-headway strategy}
As baseline holding policy, the even-headway strategy is used, which aims at equalizing the headways with respect to the leading and following vehicles. The policy is formulated as:

\begin{equation}
    \pi^{C}_{ij} =  \max\left( \frac{\hat{H}_{i+1,j} - H_{ij}}{2}, 0\right)
\end{equation}
Where $H_{ij}$ is the headway of trip $i$, or the interarrival time between trips $i$ and $i-1$ (the leading vehicle) at stop $j$ and $\hat{H}_{i+1,j}$ is the equivalent for trips $i$ and $i+1$ (the following vehicle). Because $\pi^{C}_{ij}$ is computed at the arrival of trip $i$ at stop $j$, $H_{ij}$ is simply the time elapsed since the arrival of trip $i-1$. Conversely, $H_{ij}$ requires forecasting the arrival time of trip $i+1$ at stop $j$. This can be done by integrating arrival time predictions into the methodology based on vehicle locations, or by retrieving predicted times directly from customer-facing predictions generated by the agency, as applied in \cite{cats_regularity-driven_2014,berrebi_comparing_2018, berrebi_translating_2018}. \par

The advantages of this method are its minimal data and computation requirements. The intuitive formulation also allows supervisors and drivers to more easily interpret and assess the quality of recommendations when generated by an automated system. These factors facilitate the real-time deployment of this control algorithm.\par

\subsection{Reinforcement learning}
In addition to the even headway strategy, we use a reinforcement learning approach. This section provides an overview of the methodology, but a more detailed description can be found in \cite{rodriguez_cooperative_2023}. Reinforcement learning (RL) is a data-driven method that enables the controlled agent (e.g., a transit vehicle) to learn optimal control strategies (e.g., holding) from experience in a simulation or real-world setting. As the training environment, the proposed RL approach uses a detailed simulation model that can replicate the stochasticity in vehicle travel times, dwell times, and passenger demand. As the agents explore holding strategies in this training environment, the learned control behavior is encoded in the trained decision function, which maps the observable system state to actions that maximize reward. The RL architecture requires careful design of the state parameters, which represent the decision-making factors, the set of actions considered, and the reward function, which captures the desired objectives. \par 

The action to learn is the holding time, expressed in intervals of 30 seconds (i.e., 0, 30, 60...) to limit the dimensionality of the action set. The reward function is designed to evaluate the effectiveness of holding interventions in terms of their impact on passenger waiting and riding times, and captures the benefits and costs imposed on different rider groups. Consider, for example, the waiting time of riders waiting at downstream stops of a holding action. If trip $i$ holds at stop $j$, it allows some riders at stop $j+1$ to catch trip $i$, instead of trip $i+1$ if holding was not applied. The remaining riders, however, experience an increased wait time. on board trip $i$ have longer riding times if holding is applied. To reflect these trade-offs, the reward measures the waiting and riding times of all passengers impacted in the segment between control stops. This information can be extracted from the individual passenger journey tracking in the simulator environment. The reward is then calculated using a weighted function to encourage waiting time savings and penalize added waiting and riding time. \par

The selection of decision factors embedded in the RL state aimed to balance an appropriate level of detail and computational complexity, while also considering data availability in the real world. The parameters included are as follows:
\begin{itemize}
\item     Similar to the even-headway strategy, the critical decision factors included are the headways with respect to the leading and following vehicle, represented as $H_{ij}$ and $\hat{H}_{i+1,j}$, respectively. Additionally, we consider the information from earlier trips. Because of the dynamics in transit operations, delays can propagate across consecutive trips, and headways can be interdependent (for example, bus bunching is commonly followed by a large gap). To capture this effect and enrich the operational context, we include the headway of the leading vehicle, i.e., $H_{i-1,j}$. 
    \item A vehicle can experience varying demand and travel time patterns along the route, which can influence the effectiveness of holding. To capture the spatial context, the indicator $\frac{j}{M}$ is included to represent the location of the vehicle relative to the route, where $j$ is the current control stop, and $M$ is the number of stops in the route.
    \item Another important consideration is the passenger journeys that would be impacted (positively or negatively) by holding. For this reason, we include the passenger load at stop $j$ (not including the alightings), denoted as $\mathcal{L}_{ij}$, and the number of boardings $\mathcal{B}_{ij}$. $\mathcal{L}_{ij}$ and $\mathcal{B}_{ij}$ are included separately to distinguish stops on-board passengers, which may be impacted negatively by holding (increased travel time), and passengers waiting at stops, which may be impacted positively (by a prior holding). One consideration is that demand data is typically not accessible in real-time. Hence, the estimated load $\hat{\mathcal{L}}_{ij}$ and boardings $\hat{\mathcal{B}}_{ij}$ are derived from historical averages for the time period (to account for temporal effects) and the observed headway.
\end{itemize}

The state definition with the concatenated parameters is as follows:
\begin{equation}
    s:\left[H_{ij},  \hat{H}_{i+1,j}, H_{i-1,j}, \frac{j}{M}, \hat{\mathcal{L}}_{ij}, \hat{\mathcal{B}}_{ij}\right]
\end{equation}
Where $s$ is the state. Finally, the policy that maps states to actions is determined by the learning algorithm. For this approach, the Deep Q-Networks (DQN) methodology is used \cite{mnih_human-level_2015}, which trains the parameters of a neural network-based predictor $Q_{\theta}(s,a)$ to predict the value of taking action $a$ at state $s$. After sufficient training, the final policy can be used to extract the control action as follows: 

\begin{equation}
    \pi^{RL}_{ij}=\argmax_{a}{Q_{\theta}(s,a)}
\end{equation}

Where $\pi^{RL}_{ij}$ is the RL policy and $\theta$ are the parameters of the neural network.\par 

The RL model is trained using an event-based transit simulator that reflects the observed operational conditions. The vehicle movements are modeled as arrival and departure events at each stop, with the travel time between stops sampled from an empirical distribution of observed data. The observed data could be extracted from Automatic Vehicle Location (AVL) data. Additionally, passenger journeys are modeled individually according to observed demand patterns from Automatic Passenger Count (APC) and Automated Fare Collection (AFC) data. Simulating individual passenger journeys allowed the extraction of riding and waiting times, which are inputs to the reward function.\par 

The simulator was developed in Python and was made compatible with RL training using the Gymnasium package. The neural networks were implemented with the PyTorch package. The hyperparameters that regulate the training process were calibrated for maximal performance. It is worth noting that, due to the simplified state and action formulations, the trained model can run quickly and is practical for actual deployment. 

Driver compliance impacts the effectiveness of control. To increase robustness to this driver behavior, the models were trained under scenarios with reduced compliance. Similar to \cite{phillips_quantifying_2015}, this was implemented by defining the executed holding time as a random variable uniformly distributed between 50\% and 100\% of the holding time instructed by the RL model. \par

\section{Decision Support System}\label{sec7}

The objective of the experiments is to streamline and enhance the implementation of the real-time control strategies by operations staff (e.g., supervisors and controllers). The decision support system (DSS) is designed to support real-time decisions and archive data for analytics, following the framework of \cite{maltzan_using_2015}. \par 

\begin{figure}[!ht]
    \centering
    \includegraphics[width=\linewidth]{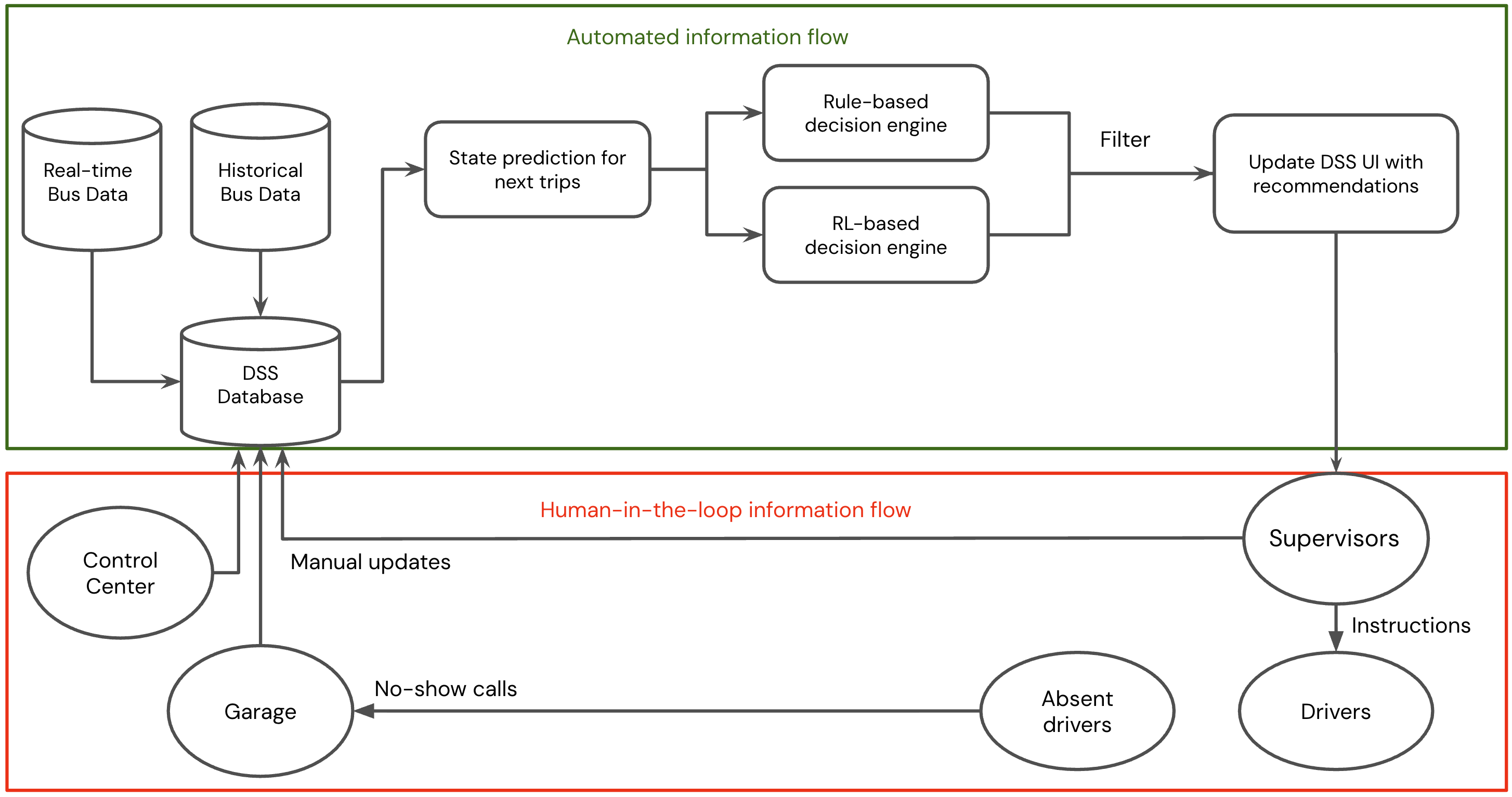}
    \caption{Information flow for the DSS}
    \label{fig:workflow}
\end{figure}

The information flow supported by the DSS is illustrated in Figure \ref{fig:workflow}, including the automated and human-in-the-loop components. These system components are described in detail in the remainder of this section.

\subsection{Automated component}
The automated information component of the DSS consists of obtaining the updated state of the bus operations from automated data sources and using the state to generate service adjustment recommendations. The required data for the DSS is a combination of offline and real-time data. The offline data includes the transit schedules from the General Transit Feed Specification (GTFS) data, which is publicly available, and historical demand data extracted from Automated Passenger Count (APC) data, which was provided by the agency. The real-time data consists of the arrival time predictions for the next trips at a given control stop of interest. \par 


The DSS automatically fuses the multiple datasets into an enriched representation of the state of the trips expected to arrive at a given control stop. First, the trips from predictions are matched to the corresponding scheduled trips in GTFS. This allows the extraction of the schedule deviation, which is a relevant input to the even-headway strategy and the RL model. It also adds details that are relevant to the user (controller or supervisor), such as the recovery time at the end of the trip, which can limit the allowable holding time. With the matched predicted and scheduled times, the predicted and scheduled headways are derived. Finally, the expected boardings for the trip, which are inputs to the RL model, can be obtained from the boarding rates based on historical APC data and the calculated headways. \par 

With up-to-date information, the DSS calls the decision models to obtain the recommended holding time. An important consideration in the design of this component is safeguarding the RL model recommendations. The quality of the RL recommendation is affected by various factors, such as missing or noisy data in the state parameters, and the presence of scenarios that were unseen during training. One technique to address this is a heuristic that limits the holding time recommended by RL to avoid excessive holding times or schedule lateness. Additionally, the holding time according to the even-headway strategy is added as an alternative recommendation. The minimal computation and data needs of the even-headway strategy make it an appropriate "fallback" model in case the RL recommendation seems inappropriate for the observed conditions.\par

\subsection{Human-in-the-loop component}
Although the automated data sources provide a comprehensive picture of the operation, they lack some details. Much of the real-time information gets manually logged by the operational staff,  whether it be on paper forms or disconnected digital systems, or verbally communicated through radio channels. Information through these alternative channels is hardly accessible but often critical to decision-making in service control. The DSS is designed as a repository of this manual service updates for real-time visibility. Updates also trigger the automated component to recalculate the state of the next trips, ensuring the holding time recommendations reflect the recent changes.\par

In addition to holding recommendations, the DSS enables two important use cases. First, it allows for manually updating service adjustments, such as holding at the terminal or mid-route stops. On the field, only terminal-based holding actions are recorded by supervisors, and usually on a paper sheet as a historical record. Given the regularity of these service adjustments, this information is not typically communicated to the control center. However, if a controller observes a flag for a trip that is running too late or too early, they may verify with the corresponding supervisor whether it was a result of a service adjustment. Overall, this information flow is prone to miscommunications. The DSS enables the supervisor to record terminal-based holding actions, making it visible in real-time to other stakeholders (such as controllers). This also applies to mid-route holding actions, which may be the initiative of supervisors or controllers, and should be recorded and visible.   
Another important use case for the human-in-the-loop communication in DSS is trip cancellations and reassignments. Bus drivers are typically managed by the administrative staff at the bus garage. When there are driver absences, garage staff must reassign their scheduled trips to either the extraboard or to a full-time driver through overtime pay. If there is no available workforce to assign a trip to, the trip is canceled. A trip cancellation is typically notified verbally to supervisors and controllers, who must adjust the service to cover the gap from the missing trip. This process has vulnerabilities, risking short or no notice on missing trips. The DSS, however, enables garage staff to record canceled trips, which can be immediately visible to decision makers. 

\subsection{User interface}
A web application was designed as an interface between field users and the DSS, thus enabling the headway management process with humans-in-the-loop. The web app combines information from automated and manual sources. It was designed to fit the workflow of users, including field supervisors, control center staff, and garage clerks, and their inputs in the development phase shaped the design. \par

\begin{figure}[!ht]
    \centering
    \includegraphics[width=\linewidth]{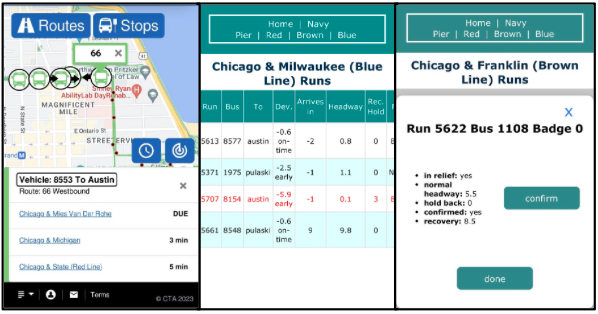}
    \caption{User toolkit: CTA BusTracker\texttrademark  application (left) and DSS web app user interface (center and right)}
    \label{fig:ui}
\end{figure}

The DSS web app can be launched on any device with an internet connection, which includes phones and tablets to fit the mobile workflow of field supervisors. Figure \ref{fig:ui} shows screenshots of the user interface (UI). In the home page, the user can select the control stop of interest from the navigation bar. The view for each control stops includes a table of the trips that are predicted to arrive at the stop, where each row is one trip. The trip row shows basic information, such as the time to arrival, the schedule deviation, and the recommended holding time. To obtain more detailed information or confirm a service action, the user can select a trip row to open a pop-up window. This window includes more details such as the scheduled recovery time at the end of the trip and the relief, which were raised by operations staff as decision factors for holding interventions. 
The window also includes a "confirm" button for the supervisor or controller to select when they have implemented the recommended holding time by communicating it to the driver. Confirming a holding action triggers the back-end to record the timestamp of the holding instruction and update the state and recommendation for the following trips. The confirmed record can help detect instances when the driver did not comply with the communicated instruction.
The window also includes a "cancel" button, that garage clerks can select in case of a trip cancellation. This can be reverted in case the trip is reassigned. Similar to the holding confirmation, changing the trip status triggers the automated component in the back-end to display recalculated headways and recommendations.

\section{Pilots}\label{sec10}
This section describes the pilot experiments in routes 81 and 66 at the Chicago Transit Authority (CTA), conducted between 2022 and 2023, and subsequently evaluates the performance impacts. The goals of the pilots are to determine the feasibility of deployment of advanced control algorithms, demonstrate effectiveness in improving reliability, and derive lessons for scalability. \par 

To evaluate the impacts of the control strategy, we compared route performance between the pilot period and the preceding weeks. The performance metrics selected comprehensively capture the impacts of interest to the various stakeholders, listed as follows:
\begin{itemize}
    \item \textbf{Passenger wait time:} The wait time is a critical indicator of service quality for riders of frequent transit services. This measure is derived from the observed headways, using the equation $\frac{\bar{H}}{2}\left(1+\left( \frac{\sigma_H}{\bar{H}}\right)^2\right)$, where $\bar{H}$ and $\sigma_H$ are the mean and standard deviation of the headways recorded at the route or stop. 
    \item \textbf{Run time variability:} The run times refer to the time elapsed between serving the first and last stop of the route for a given trip. Applying control reduces the number of trips experiencing long dwell times that result from increased passengers waiting over long headways. This effect can translate to more consistent run times, which benefits bus drivers by allowing more consistent breaks, and, in the long term, can lead to resource savings for the agency from reduced fleet requirements (which depend on run times).
    \item \textbf{Crowding:} Overcrowding can significantly impact passenger and driver experience. We therefore evaluate observed loads extracted from Automated Passenger Count (APC) data to measure how effective control strategies are in achieving a more balanced load distribution. 
    \item \textbf{Transfer times:} Riders that transfer into the route from other routes can experience reduced transfer times when reliability is improved. Transfer times are extracted from inferred trip transfers in the Origin-Destination-Transfer (ODX) data provided by the CTA. 
\end{itemize}

\subsection{Terminal control pilot}
\begin{figure}
    \centering
    \includegraphics[width=\linewidth]{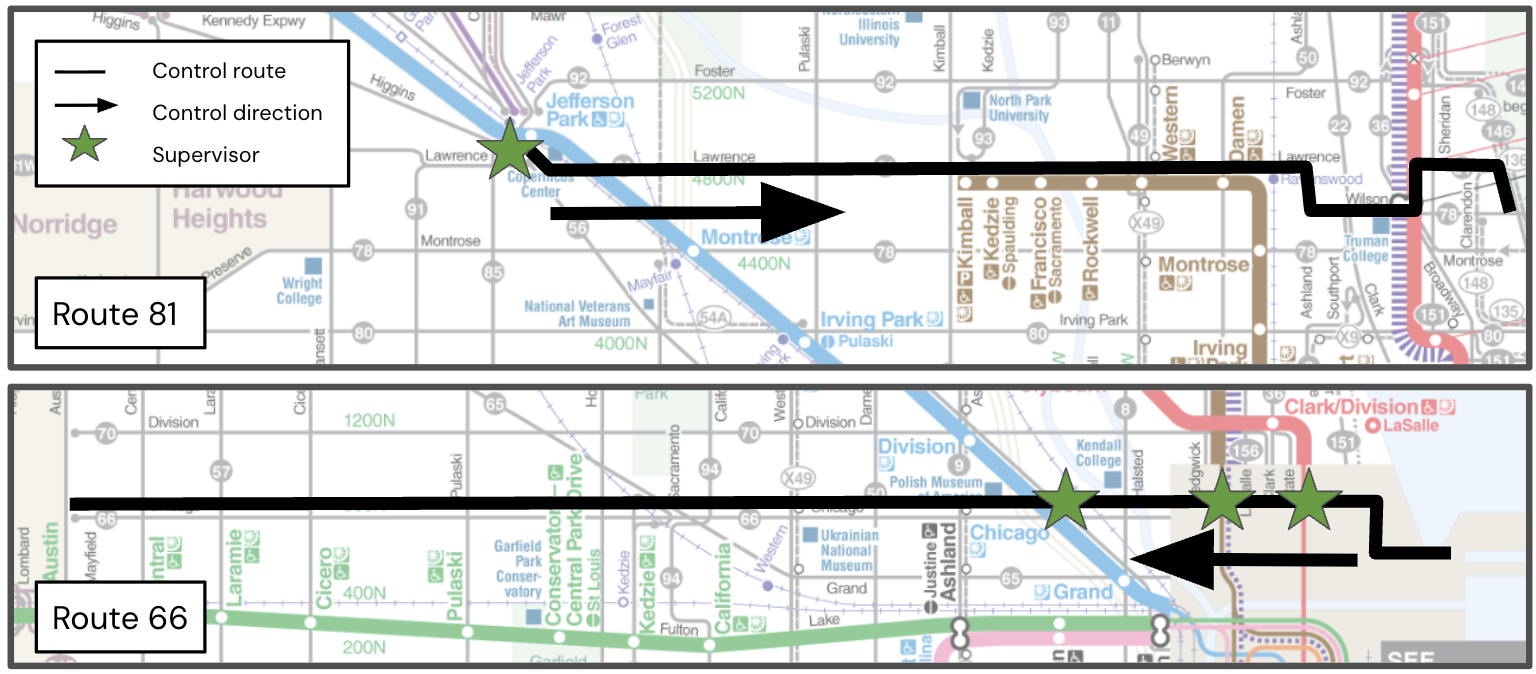}
    \caption{CTA's routes 66 and 81}
    \label{fig:route-maps}
\end{figure}

The experiment was conducted on route 81 (Figure \ref{fig:route-maps}), which runs along an east-west corridor in North Chicago and serves over 8,000 riders per weekday. 
The experiment was conducted during the week of October 17-21, 2022. During this time, the CTA, as many other agencies after the pandemic, had increased levels of driver shortages and absenteeism system-wide, which resulted in daily trip cancellations. On route 81, 15 to 25 percent of the daily scheduled trips were canceled, which made it a good candidate for control. The practice followed by supervisors was to instruct the trips following the missing trip to depart from the start terminal earlier than scheduled (a simplified version of the even-headway strategy). This pilot was focused on increasing the effectiveness of the dispatching strategy with the proposed holding control algorithm.\par

The terminal for Eastbound trips, Jefferson Park, was selected as the control point for holding. This location has a permanent supervisor managing the fleet and workforce, and is a critical transfer hub with several bus routes and the Blue Line (rail). A single supervisor is responsible for managing all routes departing the terminal, so for the period of the pilot, an additional supervisor was assigned to focus on route 81. The supervisor was instructed to communicate the holding time indicated in the DSS app to drivers at the terminal before the start of their trip, and subsequently confirm the action on the app. The garage clerks, who are the source of information on driver absences, were tasked with updating trip cancellations directly on the app. This eliminated the pre-existing need to communicate such updates over the radio or phone. Control center staff also had access to the app to monitor the interventions reported by supervisors. \par

The AM (6-9AM) and PM (3-6PM) peak periods were selected for the pilot. During the peak periods, the service had 6 to 9 minute headways. The schedule deviation thresholds, which limit the holding recommendations of the DSS models, were set as $s^{e}=4$ and $s^{l}=5$ (minutes), respectively. Additional holding control at mid-route stops was not considered.\par 

The performance of the service in the 2 weeks preceding the experiment is used as a baseline. There were no major disruptions during these periods, and the missing trip levels were similar for baseline and pilot periods.\par

\subsubsection{Compliance}
Out of the 158 trips observed during the experiment period, 80 trips (50.6\%) recorded an adjusted departure, per the DSS app records. To measure the compliance, the instructed departure communicated to the drivers was compared to the observed departure times. Following \cite{fabian_improving_2017}, trip departures within 45 seconds of the instructed time were considered compliant.\par

\begin{figure}[!ht]
  \centering
  \includegraphics[width=\textwidth]{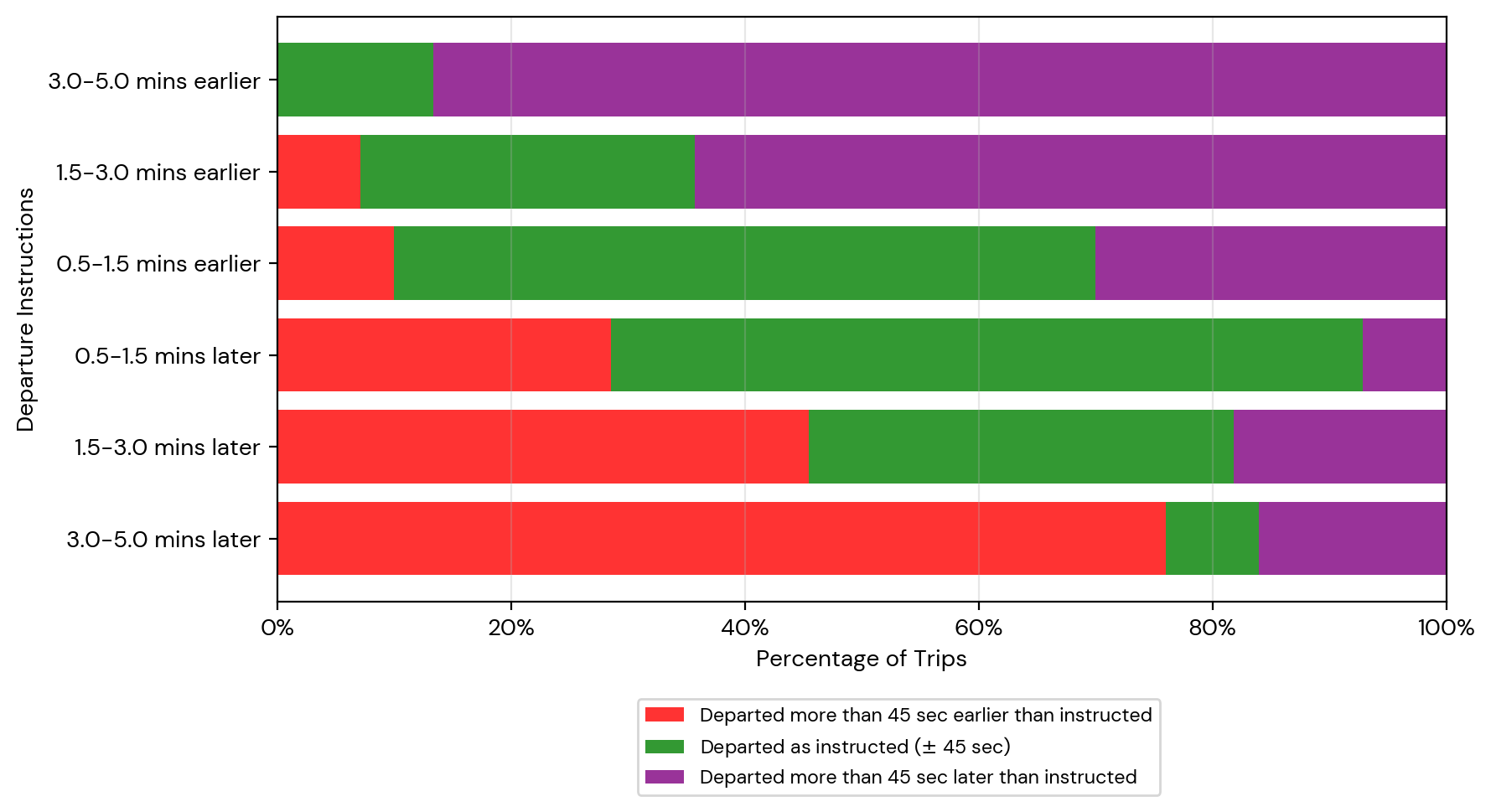}
  \caption{Compliance levels as a function of the instructed adjustment relative to the scheduled time}\label{fig:compliance_breakdown}
\end{figure}

Using this method, 28 of the 80 instructed departures were identified as compliant, indicating a compliance rate of 35\%.
To investigate the patterns of compliance further, Figure \ref{fig:compliance_breakdown} shows a breakdown of compliance by how far the instructed time deviated from the scheduled time. It is noteworthy that the further the instructed departure is from the scheduled time, the lower the compliance rate is. Furthermore, non-compliant trips tend to depart closer to the scheduled time. For example, for trips instructed to depart 3-5 minutes earlier than scheduled, all non-compliant trips departed later than instructed, or closer to the scheduled time. This is consistent with the findings of  \cite{fabian_improving_2017} in a similar experiment on a rail line in Boston and in part can be explained by the agency's focus on schedule punctuality. \par 

\subsubsection{Overall performance}
Table \ref{tab:pilot_metrics} compares the performance of the baseline and pilot periods in the AM and PM peak. The table includes the average number of trips per day to account for service levels. The wait time, 90th percentile run times, and 90th percentile crowding are also reported.\par 

In the AM period, the performance is improved across all metrics. The wait time was reduced by 8.7\%, saving riders an average of 0.6 minutes of waiting time. The scenarios had a similar number of daily trips, which highlights the effectiveness of the pilot strategy. In the PM period, the benefits were more substantial. The regularity improvements translated to 1.6 minutes in average wait time savings per rider. The larger improvement margin observed in the PM period can be attributed to several factors. The conditions during the PM period are more challenging, making the service quality more vulnerable to missing trips. Based on observations before the experiment, the supervisor responsible for the PM period was less effective in managing departures than the AM period supervisor. \par

Overall, the analysis shows that despite the lower compliance, the performance benefits from control are significant. This finding is consistent with a simulation-based study \cite{phillips_quantifying_2015} that showed performance improvements in service regularity can be gained even with lower compliance levels. 

\begin{table}
\caption{Pilot metrics}
\label{tab:pilot_metrics}
{\tiny
\begin{tabular}{lccccccccc}
\toprule
stage & \multicolumn{3}{c}{81 AM} & \multicolumn{3}{c}{81 PM} & \multicolumn{3}{c}{66} \\
pilot & base & pilot & change (\%) & base & pilot & change (\%) & base & pilot & change (\%) \\
metric &  &  &  &  &  &  &  &  &  \\
\midrule
\textbf{avg trips per day} & 20.7 & 21.2 & 0.5 (2.4\%) & 20.9 & 22.2 & 1.3 (6.2\%) & 43.3 & 41.7 & -1.6 (-3.7\%) \\
\textbf{wait time (mins)} & 7.25 & 6.62 & -0.6 (-8.7\%) & 8.28 & 6.63 & -1.6 (-19.9\%) & 5.13 & 4.83 & -0.3 (-5.8\%) \\
\textbf{90th run time (mins)}& 59.9 & 57.1 & -2.8 (-4.7\%) & 61.8 & 60.7 & -1.1 (-1.8\%) & 71.9 & 71.2 & -0.7 (-1.0\%) \\
\textbf{90th load}& 56.0 & 54.0 & -2.0 (-3.6\%) & 65.0 & 54.9 & -10.1 (-15.5\%) & 60.0 & 56.7 & -3.3 (-5.5\%) \\
\bottomrule
\end{tabular}
}
\end{table}

\subsubsection{Performance by service levels}
\begin{figure}
    \centering
    \includegraphics[width=\linewidth]{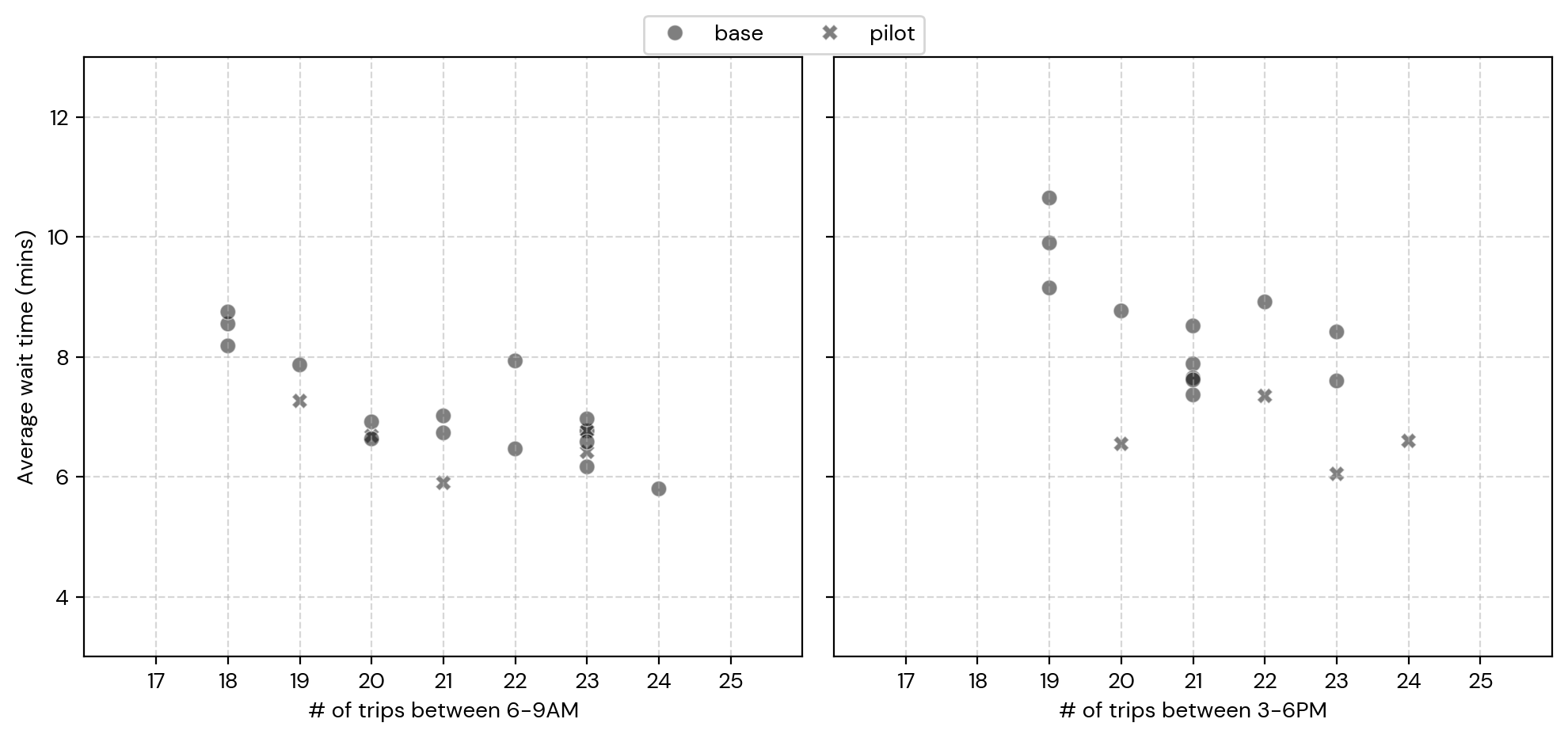}
    \caption{Wait time performance by the daily number of trips observed in route 81  in the AM (left) and PM (right) peak hours, labeled by baseline and pilot}
    \label{fig:wt_vs_service}
\end{figure}

The number of missing trips due to operator absenteeism varied day-to-day. This makes it challenging to isolate the effect of the proposed holding algorithm from the effect of service levels. To analyze the performance difference at different service levels, Figure \ref{fig:wt_vs_service} shows the wait time by the number of trips observed in the AM and PM evaluation periods, where each point represents one day.\par

The range of service levels observed in the AM period in the weeks of collected data ranges from 17 to 24 trips, and 19 to 24 trips in the PM period. As expected, the wait times decrease with an increased number of trips. However, the pilot strategy reduced wait times for days with similar service levels, and performed as well as days with more trips in the baseline period. In the AM peak, the pilot's wait time in the day with 19 trips was comparable to the baseline's wait time with 20 trips. Similarly, the pilot's performance with 21 trips was comparable to the baseline performance with 23-24 trips. It is worth noting that for the 23-trip day, the pilot did not improve performance significantly compared to the baseline. It may be the case that for higher service levels, the trip headways are more likely to fall out of balance downstream and require some additional corrective actions at intermediate stops. The performance improvements are more pronounced in the PM peak. Across all observed days, the wait time during the pilot was consistently between 6 and 7 minutes, whereas the wait time in the baseline weeks was between 7 and 9 minutes.\par

As the results show, applying the holding strategy can improve performance as much as increasing service does. This finding shows that control can be a cost-effective solution to improve service for riders. Additionally, the consistent performance for different service levels highlights the added robustness of the proposed strategy to service level variations. \par

\subsection{En-route control pilot}
\subsubsection{Pilot description}

A second pilot was conducted on route 66 (Figure \ref{fig:route-maps}) between November 6 and 17, 2023. The route runs along a major east-west corridor that serves various residential and employment areas in the inner core region of Chicago, and connects to major bus and rail lines. At the time of testing, the route served circa 15,000 passenger rides every weekday.\par 

Based on preliminary analysis, it was found that delays and headway variability issues in peak hours originated mostly in the early portion of the westbound direction. This area is a dense activity center, with congested traffic conditions affecting travel times, and high transit ridership increasing dwell times. Additionally, a portion of the fleet is e-buses, which require charging times at the terminal that can delay the trip start. Prior to the start of the experiment, no control strategies were applied. To address these issues, it was proposed to pilot the holding strategy at three control stops. The stops selected, shown in Figure \ref{fig:route-maps}, are transfer stops to the Red, Brown, and Blue rail lines, respectively. Selecting these stops minimized the need for driver re-training, since these are locations at which bus drivers must hold if they arrive earlier than scheduled (referred to as timepoints). Additionally, the higher demand at these stops makes most buses stop and dwell, making it easier to apply holding by simply extending the dwell time. It is also perceived more favorably by riders, since the holding time allows more riders to board. Preliminary analysis found that service regularity at the terminal was not affected, and thus it was not used as a control point.\par

The CTA allocated two supervisors to apply the holding strategy. The supervisors were equipped with a mobile device to access the DSS app and the public bus tracking map (shown in Figure \ref{fig:ui}). The supervisors were instructed to communicate the app-based holding instruction to drivers at the stop and confirm the action in the app. The supervisors at CTA do not have a direct communication line with bus drivers; therefore, the protocol was for the supervisors to physically meet the drivers at the stop. Each supervisor managed one stop at a time and drove a short distance to another control stop if needed. A research assistant from Northeastern University was remotely monitoring the functionality and use of the app, and was available by phone or text for troubleshooting. \par

The schedule deviation threshold for late trips was determined as $s^{l}=5$ minutes. $s^{e}$ was not relevant since typically trips do not arrive early at the mid-route. The holding strategy was applied during the weekday PM rush hours (2-6PM).  At this time of day, most westbound trips are short-turned several stops before the terminal to meet the demand of the earlier portion of the route. Therefore, only the stops in this route segment were the focus of the evaluation. The two weeks preceding the experiment were used as a baseline period for comparison.\par

\subsubsection{Compliance}
Over the two weeks of the pilot, the app logs recorded 174 holding instructions. Similar to route 81, we measure how many of these holding actions were followed by the drivers. A heuristic is employed to identify compliant trips based on the deviation between the observed dwell time and the theoretical dwell time. The theoretical dwell time is calculated from the trip's observed boarding and alighting times (based on APC data) and historical average boarding and alighting times, similar to the methodology used in  \cite{sindzingre_detecting_2019}.  Trips with instructed holding and dwell time deviations exceeding the 90th percentile of dwell time deviations for non-instructed trips are considered compliant.\par 

This method identified 99 compliant trips, corresponding to a compliance rate of 56.9\%. The higher compliance relative to the route 81 pilot can be attributed to the presence of a supervisor to enforce the holding time. It is also worth noting that another factor impacting the holding compliance is that some control stops are located immediately before a signalized intersection, which can encourage drivers to leave before the instructed holding to avoid being stuck for another signaling cycle.\par

\subsubsection{Overall performance}
Table \ref{tab:pilot_metrics} compares performance for various metrics. The results indicate improvement across the board, but with more muted improvements on the 90th percentile run times. The improvements in service regularity translated to an average of 0.3 minutes of wait time saved per rider, or 5.7\% of the total. More regularity also reduced overcrowded buses; the 90th percentile of observed bus loads reduced from 60.0 to 56.7 riders. Considering that this is a route with crushing loads, these improvements could translate to fewer denied boardings. \par

The improvement showcases the effectiveness of the strategy despite the limited scale of holding control, with holding being applied only at two stops located in the early part of the route. Furthermore, there were 1.6 fewer trips per day (on average) in the pilot compared to the baseline, highlighting the strategy's robustness to missing trips.\par 

\subsubsection{Wait time}
\begin{figure}
    \centering
    \includegraphics[width=\linewidth]{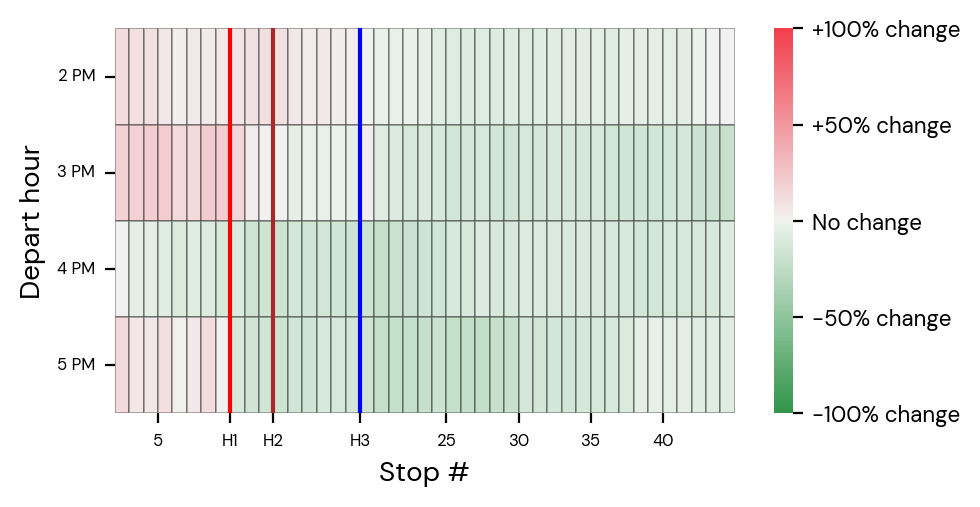}
    \caption{Percentage change in wait time from baseline to pilot period, by hour (based on the terminal departure time). Holding stops labeled as H1, H2, H3, respectively.}
    \label{fig:wait_time_change}
\end{figure}

The spatial performance of wait time is analyzed to evaluate whether holding improves performance downstream. Figure \ref{fig:wait_time_change} shows the percentage change in wait time from baseline to experiment period for every stop. Additionally, to isolate the effect of temporal conditions within the PM peak, the performance is also separated by the hour of departure. The results show worse or equal performance before the holding stops (labeled in the chart as H1, H2, H3), and a significant improvement following the holding stops. The performance also aligns with the locations chosen for holding within the period. Between 2-4PM, most interventions were focused on H2 and H3, and between 4-6PM holding was applied at H1 and H3, and this is well validated by the results. Importantly, the wait time savings were sustained for the remainder of the route, which is significant considering how long the route is (6 miles or 47 stops). This validates the choice of holding stops, showing that when corrective strategies are applied at the source of delays, this can have a compounding benefit on the rest of the route.

\subsubsection{Transfer times}
Given the high ridership at the transfer stops of route 66, the effect on transfer times is evaluated. Data on transfer journeys were obtained from the CTA's Origin-Destination-Transfer (ODX) datasets. The ODX data is generated from a destination and transfer inference process applied to Automated Fare Collection (AFC) data, based on the methods detailed in \cite{gordon_automated_2013}. The data only includes transfers between bus routes. For significance, stops with transfer volumes of more than 5 passengers per hour were considered. The process resulted in a sample size of 3,154 inferred transfers across six stops in the baseline and pilot periods. \par

\begin{figure}
    \centering
    \includegraphics[width=\linewidth]{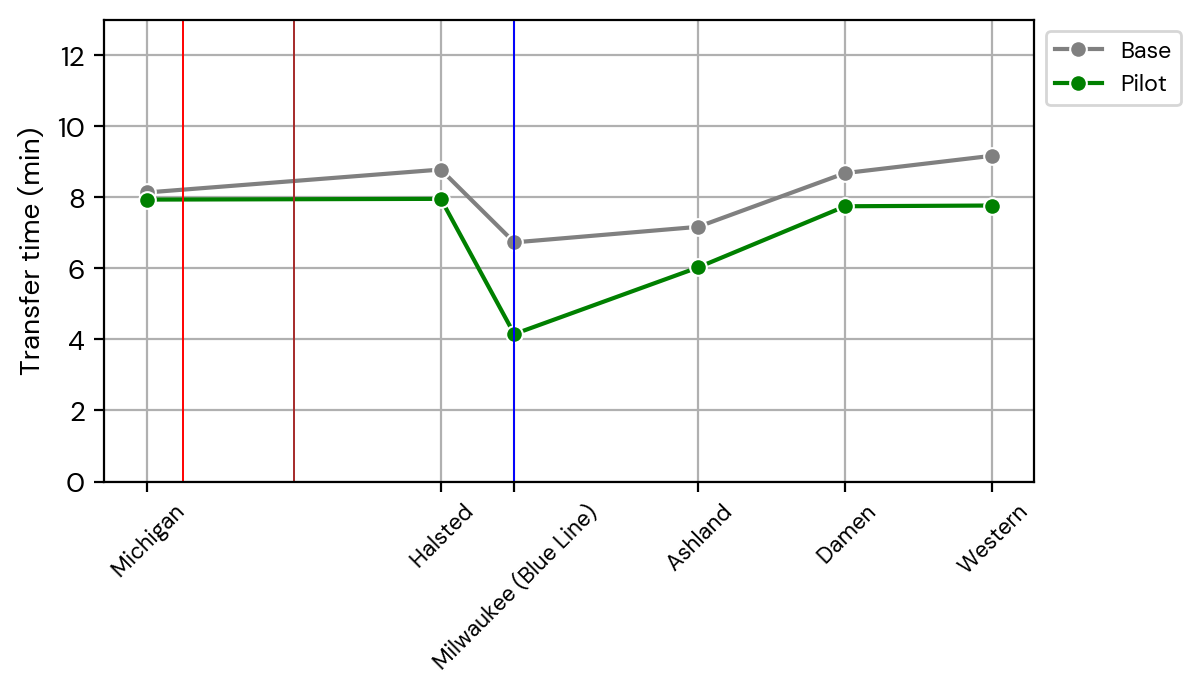}
    \caption{Transfer times at key transfer stops along route 66}
    \label{fig:transfer-times}
\end{figure}

Figure \ref{fig:transfer-times} compares the 75th percentile transfer times.  Overall, the transfer times vary along the route, which is likely due to the different service frequencies of the routes that riders transfer from. Following the trends in wait time, the 75th percentile transfer times are similar for both scenarios before the first holding point, and lower for the pilot after the holding points, indicating an impact from the strategy. This reduction in the 75th percentile is expected, since increased service regularity results in less extreme headways, i.e., large gaps.\par

This improvement is significant considering that riders transferring from other routes are unable to time their arrival with journey planning apps, and thus are more susceptible to longer headways. \par

\subsubsection{Short turning}
Aside from applying holding, supervisors were encouraged to use the monitoring capabilities of the DSS app to manage service more broadly, and if appropriate, implement other service adjustments they are trained in. The objective of this exercise was a proof-of-concept on how well the DSS would generalize to the custom workflows of supervisors on different routes. Short-turning trips coming in the eastbound (opposite) direction were the tested strategy. The predictions available in the DSS app and bus locations in the tracker app would alert supervisors about eastbound trips with extreme delays, which were at risk of propagating into the following westbound trip. In some of those cases, the supervisors instructed those buses to short-turn at a suitable location a few stops before the terminal, allowing the trip to get back on schedule. Riders on-board (typically few) were asked to alight and wait for the next trip, which would not be long given the high-frequency during this period.\par

\begin{figure}
    \centering
    \includegraphics[width=0.8\linewidth]{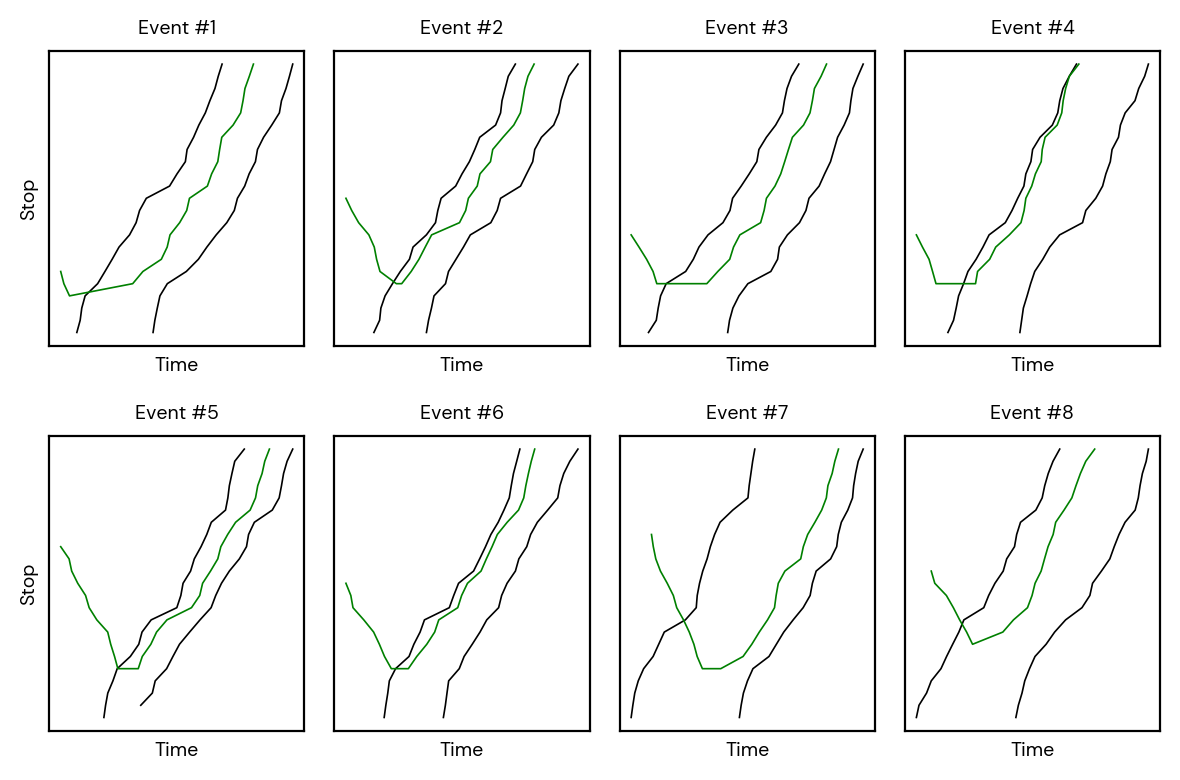}
    \caption{Trip trajectories for short-turning events}
    \label{fig:short_turns}
\end{figure}

This strategy was deployed eight times, which is insufficient to allow a systematic analysis of the benefits. For a qualitative assessment, Figure \ref{fig:short_turns} shows the trajectories for the short-turning events, including the short-turned trip and its leading and following trips. For the most part, the short-turned trips were evenly spaced between leading and following trips, indicating an effective implementation. Interestingly, the trajectories in events 2, 3, and 5 suggest that en-route holding was applied to the short-turned trip to correct the uneven spacing upon departure. This showcases the potential of a combined strategy, validating a simulation-based study that integrated both strategies into the methodology \cite{tang_robust_2024}. \par

\section{Implications for Scalability}\label{sec12}
The analysis highlighted that headway control can have substantial benefits in service quality. To understand the implications for the scalability of this implementation, this section provides lessons learned from the data and observations. 

\subsection{Resource requirements}
The holding strategy was resource-efficient and quick to deploy. The web-based application ran on existing smartphones, and the training involved only a few supervisors, as drivers were already trained to follow the types of strategies tested. In both pilots, dedicated supervisors were needed, which represented additional labor costs. However, this would not be necessary in the long term, as supervisors become better trained to use the web application. The route 81 pilots showed that the benefits of control can be as significant as adding service. Having an additional supervisor can be a fraction of the added operational cost of more buses and drivers, making it an efficient investment for transit agencies.\par 

For routes without an active supervisor, there is potential in using automated driver instructions via the on-board digital device, as is implemented in Stockholm \cite{cats_regularity-driven_2014} and Chile \cite{lizana_bus_2014}.  The radio-based communication channel between the control center and drivers, which is typically in place at agencies, can also be used for control instructions. This would add some upfront costs, including driver re-training.\par 


\subsection{Digital infrastructure}
A benefit of the DSS development is that it fills an information and communication gap for field supervisors. Regarding monitoring technology, the tools currently used by supervisors are more limited than those available to the control center staff. The existing CTA supervisors' tools are not easy to navigate for monitoring a single route, do not flag trips in need of corrections, and do not allow the real-time recording of these interventions for other staff to be aware of. Despite supervisors having similar responsibilities as controllers, the information access is different. This is also the case in the communication channels; direct communication to drivers is only allowed to controllers, which limits the supervisor's ability to implement interventions. The pilot also showed that equipping field staff with adequate information tools, like the DSS, can enable more complex corrective strategies. This was evidenced by the effective implementation of short-turning, which was not previously considered in the planning of the pilot.

\subsection{Driver behavior}
\begin{figure}
    \centering
    \includegraphics[width=0.6\linewidth]{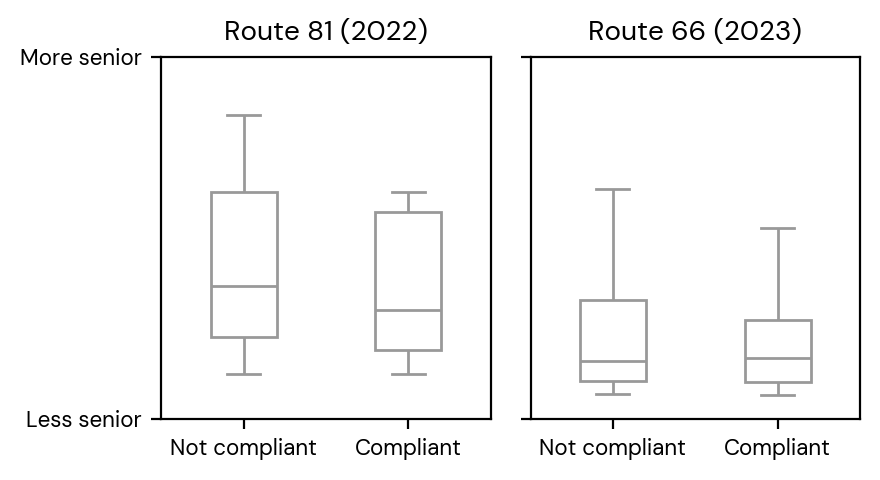}
    \caption{Distribution of seniority by compliance}
    \label{fig:seniority-compliance}
\end{figure}

The pilots revealed the presence of driver non-compliance. The route 81 pilot showed lower compliance in recommended departure times that deviate further from schedule, as shown in Figure \ref{fig:seniority-compliance}. Despite communicating to drivers that CTA's established policies of enforcing punctuality would be paused for the duration of the pilot, it is possible that drivers resisted breaking legacy protocols. Additionally, the perception among some drivers is that departing a trip later than scheduled can cut into their break time at the end of the trip. As the evaluation showed, more regularity can result in more consistent run times, making break times more consistent. Training campaigns to communicate these benefits could make drivers more receptive. Another strategy to address concerns over break times is to relax the scheduled trip start and end times to allow guaranteed breaks. The headway control algorithms could accommodate this constraint in the back-end.\par

Seniority could have an influence on compliance. Less experienced drivers may follow directions more than their experienced counterparts. More experienced drivers may be more confident in self-regulating in response to delays. To investigate seniority patterns in compliance, we compare the seniority of drivers operating compliant and non-compliant trips. Figure \ref{fig:seniority-compliance} shows the seniority distributions for non-compliant and compliant trips in both pilots. It can be observed that the seniority is higher for drivers of non-compliant trips, and this is consistent across both pilots. This finding is consistent with previous driver surveys, which found that senior drivers were less likely to use real-time headway control tools \cite{martinez-estupinan_understanding_2022}.  There may be selection bias, as junior drivers’ performance can be more affected by service disruptions or delays, necessitating control actions to adjust the headways. For system-wide deployment, agencies could benefit from better training and incentives to increase compliance from drivers, particularly those most experienced.\par

\subsection{Selection of control strategies}
The analysis showed that departure adjustments can be enough to increase service quality, as demonstrated on route 81. However, the performance improvement margin was smaller in days with increased levels of service. This indicates that having additional holding strategies en route may be needed to help correct for headway disturbances formed downstream. This issue of en-route disruptions was observed in route 66, where, despite regular departures at the terminal, the service quickly deteriorated downstream due to uncertainty. In this case, there was an increased need for en-route holding. This shows that there may not be a one-size-fits-all solution across all routes, and operational staff must therefore tailor the strategy to the challenges faced by the specific route. 

The results indicated that although short-turning actions were performed manually without decision support technology, they may have been beneficial. To scale more complex strategies like short-turning effectively, understanding the performance of observed interventions is critical, along with establishing a feedback loop. The DSS is a first step in enabling this feedback loop, since it archives interventions reported by supervisors for analytics. For methodological challenges, advanced methods can be adapted for system-wide deployment. The RL recommendation models can add stop-skipping and short-turning to the set of possible actions, as proposed in recent RL-based methods \cite{rodriguez_cooperative_2023, tang_robust_2024}, and be iteratively trained based on incoming real-world performance data. Strategies may also leverage bus arrival predictions with predictive algorithms \cite{berrebi_comparing_2018}. Also, balancing the complex costs associated with short-turning (i.e., inconveniencing passengers onboard) may require more detailed optimization algorithms as those proposed in \cite{gkiotsalitis_cost-minimization_2019}. \par 

\section{Conclusion}\label{sec13}
This study presents the design and deployment of two pilots to test advanced holding control algorithms, bringing theory to practice. The strategy, high-frequency bus routes, is based on a reinforcement learning model for holding control.  Given the black-box nature of methods like reinforcement learning, guardrails were placed on the recommendations based on service-related constraints, such as maximum allowed schedule deviation. Additionally, recommendations were also generated according to the even-headway strategy, which has been well-tested in other cities and has lighter data requirements. These techniques enabled robust deployment in two pilots, which were conducted on high-ridership routes in Chicago to test terminal-based and en-route holding, respectively. Across both pilots, the system was adopted by operations staff to implement hundreds of interventions. A comprehensive evaluation of the impacts, including metrics relevant to riders, drivers, and agencies, revealed significant improvements in service quality. Recommendations are provided for scalable implementations of the strategies.\par

Extensions of this work can leverage the collected performance data from the pilots to retrain the RL models and improve the robustness of recommendations. Integrating back the real-world feedback can also reduce the difference between simulated and real-world conditions (e.g., in driver compliance) and provide information on unobserved effects. The control actions supported by the models can also be extended to include short-turning, which was applied during the pilots by the supervisors based on their own intuition and training. Short-turning is a complex strategy, and its effectiveness can depend on a variety of conditions that may be difficult to simulate a priori (e.g., the time needed for offboarding riders and turning around the bus). Therefore, it is critical to consider a human-in-the-loop approach that leverages observed real-world examples implemented by supervisors and integrates human expertise.\par

It would be beneficial to simulate driver compliance behaviors, which were assumed to be a simplified random process. The question of how compliance levels differ for different strategies, such as speed control, remains. Although driver compliance can be addressed with more training and enforcement, it is important to understand the motivations for compliance behavior, as it can reveal faults in the technology or limitations in its implementation. For instance, a driver may be reluctant to hold at a stop if it blocks vehicle traffic, which would warrant reconsidering the application of holding at that stop. Furthermore, the design of effective incentive structures requires further investigation. \par 

\backmatter

\bmhead{Acknowledgements}



This material is based upon work supported by the U.S. Department of Energy’s Office of Energy Efficiency and Renewable Energy (EERE) under the Vehicle Technology Program Award Number DE-EE0009211. The views expressed herein do not necessarily represent the views of the U.S. Department of Energy or the United States Government. Additionally, the authors would like to acknowledge the generous support from the Chicago Transit Authority in the preparation and implementation of the experiments.

\section*{Declarations}
Not applicable.

\bibliography{references}

\begin{thebibliography}{27}
\providecommand{\natexlab}[1]{#1}
\providecommand{\url}[1]{{#1}}
\providecommand{\urlprefix}{URL }
\providecommand{\doi}[1]{\url{https://doi.org/#1}}
\providecommand{\eprint}[2][]{\url{#2}}
 \bibcommenthead

\bibitem[{Abkowitz and Lepofsky(1990)}]{abkowitz_implementing_1990}
Abkowitz MD, Lepofsky M (1990) Implementing {Headway}‐{Based} {Reliability} {Control} on {Transit} {Routes}. Journal of Transportation Engineering 116(1):49--63. \doi{10.1061/(ASCE)0733-947X(1990)116:1(49)}, \urlprefix\url{https://ascelibrary.org/doi/10.1061/%28ASCE%290733-947X%281990%29116%3A1%2849%29}, publisher: American Society of Civil Engineers

\bibitem[{Argote-Cabanero et~al.(2015)Argote-Cabanero, Daganzo, and Lynn}]{argote-cabanero_dynamic_2015}
Argote-Cabanero J, Daganzo CF, Lynn JW (2015) Dynamic control of complex transit systems. Transportation Research Part B: Methodological 81:146--160. \doi{10.1016/j.trb.2015.09.003}, \urlprefix\url{https://linkinghub.elsevier.com/retrieve/pii/S0191261515001976}

\bibitem[{Bartholdi~III and Eisenstein(2012)}]{bartholdi_iii_self-coordinating_2012}
Bartholdi~III JJ, Eisenstein DD (2012) A self-coördinating bus route to resist bus bunching. Transportation Research Part B: Methodological 46(4):481--491. Publisher: Elsevier

\bibitem[{Berrebi et~al.(2018{\natexlab{a}})Berrebi, Crudden, and Watkins}]{berrebi_translating_2018}
Berrebi SJ, Crudden SO, Watkins KE (2018{\natexlab{a}}) Translating research to practice: {Implementing} real-time control on high-frequency transit routes. Transportation Research Part A: Policy and Practice 111:213--226. \doi{10.1016/j.tra.2018.03.008}, \urlprefix\url{https://linkinghub.elsevier.com/retrieve/pii/S0965856417312685}

\bibitem[{Berrebi et~al.(2018{\natexlab{b}})Berrebi, Hans, Chiabaut, Laval, Leclercq, and Watkins}]{berrebi_comparing_2018}
Berrebi SJ, Hans E, Chiabaut N, et~al (2018{\natexlab{b}}) Comparing bus holding methods with and without real-time predictions. Transportation Research Part C: Emerging Technologies 87:197--211. \doi{10.1016/j.trc.2017.07.012}, \urlprefix\url{https://linkinghub.elsevier.com/retrieve/pii/S0968090X17302000}

\bibitem[{Cats(2014)}]{cats_regularity-driven_2014}
Cats O (2014) Regularity-driven bus operation: {Principles}, implementation and business models. Transport Policy 36:223--230. \doi{10.1016/j.tranpol.2014.09.002}, \urlprefix\url{https://www.sciencedirect.com/science/article/pii/S0967070X14001917}

\bibitem[{Chen et~al.(2024)Chen, Chen, Wang, and Fang}]{chen_bus_2024}
Chen W, Chen Y, Wang Y, et~al (2024) Bus {Drivers}’ {Behavioral} {Intention} to {Comply} with {Real}-{Time} {Control} {Instructions}: {An} {Empirical} {Study} from {China}. Sustainability 16(9):3623. \doi{10.3390/su16093623}, \urlprefix\url{https://www.mdpi.com/2071-1050/16/9/3623}, number: 9 Publisher: Multidisciplinary Digital Publishing Institute

\bibitem[{Englisher(1984)}]{englisher_minneapolis-st_1984}
Englisher LS (1984) Minneapolis-{St}. {Paul} transit service reliability demonstration. {Final} report. Tech. Rep. UMTA-MA-06-0049-83-8, Urban Mass Transportation Administration, Washington, D.C.

\bibitem[{Fabian(2017)}]{fabian_improving_2017}
Fabian JJ (2017) Improving high-frequency transit reliability : a case study of the {MBTA} {Green} {Line} through simulation and field experiments of real-time control strategies. Thesis, Massachusetts Institute of Technology, \urlprefix\url{https://dspace.mit.edu/handle/1721.1/111426}, accepted: 2017-09-15T15:33:40Z Journal Abbreviation: MBTA Green Line through simulation and field experiments of real-time control strategies

\bibitem[{Fabian et~al.(2018)Fabian, Sánchez-Martínez, and Attanucci}]{fabian_improving_2018}
Fabian JJ, Sánchez-Martínez GE, Attanucci JP (2018) Improving {High}-{Frequency} {Transit} {Performance} through {Headway}-{Based} {Dispatching}: {Development} and {Implementation} of a {Real}-{Time} {Decision}-{Support} {System} on a {Multi}-{Branch} {Light} {Rail} {Line}. Transportation Research Record 2672(8):363--373. \doi{10.1177/0361198118794534}, \urlprefix\url{https://doi.org/10.1177/0361198118794534}, publisher: SAGE Publications Inc

\bibitem[{Gkiotsalitis et~al.(2019)Gkiotsalitis, Wu, and Cats}]{gkiotsalitis_cost-minimization_2019}
Gkiotsalitis K, Wu Z, Cats O (2019) A cost-minimization model for bus fleet allocation featuring the tactical generation of short-turning and interlining options. Transportation Research Part C: Emerging Technologies 98:14--36. \doi{10.1016/j.trc.2018.11.007}, \urlprefix\url{https://www.sciencedirect.com/science/article/pii/S0968090X18308040}

\bibitem[{Gordon et~al.(2013)Gordon, Koutsopoulos, Wilson, and Attanucci}]{gordon_automated_2013}
Gordon JB, Koutsopoulos HN, Wilson NHM, et~al (2013) Automated {Inference} of {Linked} {Transit} {Journeys} in {London} {Using} {Fare}-{Transaction} and {Vehicle} {Location} {Data}. Transportation Research Record 2343(1):17--24. \doi{10.3141/2343-03}, \urlprefix\url{https://doi.org/10.3141/2343-03}, publisher: SAGE Publications Inc

\bibitem[{Lizana et~al.(2014)Lizana, Muñoz, Giesen, and Delgado}]{lizana_bus_2014}
Lizana P, Muñoz JC, Giesen R, et~al (2014) Bus {Control} {Strategy} {Application}: {Case} {Study} of {Santiago} {Transit} {System}. Procedia Computer Science 32:397--404. \doi{10.1016/j.procs.2014.05.440}, \urlprefix\url{https://linkinghub.elsevier.com/retrieve/pii/S1877050914006401}

\bibitem[{Maltzan(2015)}]{maltzan_using_2015}
Maltzan DDW (2015) Using real-time data to improve reliability on high-frequency transit services. Thesis, Massachusetts Institute of Technology, \urlprefix\url{https://dspace.mit.edu/handle/1721.1/99541}, accepted: 2015-10-30T18:34:05Z

\bibitem[{Martínez-Estupiñan et~al.(2022)Martínez-Estupiñan, Delgado, Muñoz, and Watkins}]{martinez-estupinan_understanding_2022}
Martínez-Estupiñan Y, Delgado F, Muñoz JC, et~al (2022) Understanding what elements influence a bus driver to use headway regularity tools: case study of {Santiago} public transit system. Transportmetrica A: Transport Science pp 1--34. \doi{10.1080/23249935.2022.2025950}, \urlprefix\url{https://www.tandfonline.com/doi/full/10.1080/23249935.2022.2025950}

\bibitem[{Martínez-Estupiñan et~al.(2023)Martínez-Estupiñan, Delgado, Muñoz, and Watkins}]{martinez-estupinan_improving_2023}
Martínez-Estupiñan Y, Delgado F, Muñoz JC, et~al (2023) Improving the performance of headway control tools by using individual driving speed data. Transportation Research Part A: Policy and Practice 174:103761. \doi{10.1016/j.tra.2023.103761}, \urlprefix\url{https://www.sciencedirect.com/science/article/pii/S0965856423001817}

\bibitem[{Mnih et~al.(2015)Mnih, Kavukcuoglu, Silver, Rusu, Veness, Bellemare, Graves, Riedmiller, Fidjeland, Ostrovski, Petersen, Beattie, Sadik, Antonoglou, King, Kumaran, Wierstra, Legg, and Hassabis}]{mnih_human-level_2015}
Mnih V, Kavukcuoglu K, Silver D, et~al (2015) Human-level control through deep reinforcement learning. Nature 518(7540):529--533. \doi{10.1038/nature14236}, \urlprefix\url{http://www.nature.com/articles/nature14236}

\bibitem[{Pangilinan et~al.(2008)Pangilinan, Wilson, and Moore}]{pangilinan_bus_2008}
Pangilinan C, Wilson N, Moore A (2008) Bus {Supervision} {Deployment} {Strategies} and {Use} of {Real}-{Time} {Automatic} {Vehicle} {Location} for {Improved} {Bus} {Service} {Reliability}. Transportation Research Record 2063(1):28--33. \doi{10.3141/2063-04}, \urlprefix\url{https://doi.org/10.3141/2063-04}, publisher: SAGE Publications Inc

\bibitem[{Phillips et~al.(2015)Phillips, del Rio, Muñoz, Delgado, and Giesen}]{phillips_quantifying_2015}
Phillips W, del Rio A, Muñoz JC, et~al (2015) Quantifying the effects of driver non-compliance and communication system failure in the performance of real-time bus control strategies. Transportation Research Part A: Policy and Practice 78:463--472. \doi{10.1016/j.tra.2015.06.005}, \urlprefix\url{https://www.sciencedirect.com/science/article/pii/S0965856415001639}

\bibitem[{Rodriguez et~al.(2023)Rodriguez, Koutsopoulos, Wang, and Zhao}]{rodriguez_cooperative_2023}
Rodriguez J, Koutsopoulos HN, Wang S, et~al (2023) Cooperative bus holding and stop-skipping: {A} deep reinforcement learning framework. Transportation Research Part C: Emerging Technologies 155:104308. \doi{10.1016/j.trc.2023.104308}, \urlprefix\url{https://www.sciencedirect.com/science/article/pii/S0968090X23002978}

\bibitem[{Sindzingre(2019)}]{sindzingre_detecting_2019}
Sindzingre M (2019) Detecting and quantifying bus operation impedance : the balance between reliability and speed. Thesis, Massachusetts Institute of Technology, \urlprefix\url{https://dspace.mit.edu/handle/1721.1/123901}, accepted: 2020-02-28T20:50:08Z

\bibitem[{Soza-Parra et~al.(2019)Soza-Parra, Cats, Carney, and Vanderwaart}]{soza-parra_lessons_2019}
Soza-Parra J, Cats O, Carney Y, et~al (2019) Lessons and {Evaluation} of a {Headway} {Control} {Experiment} in {Washington}, {D}.{C}. Transportation Research Record 2673(8):430--438. \doi{10.1177/0361198119845369}, \urlprefix\url{https://doi.org/10.1177/0361198119845369}, publisher: SAGE Publications Inc

\bibitem[{Strathman et~al.(2001)Strathman, Kimpel, Dueker, Gerhart, Turner, Griffin, and Callas}]{strathman_bus_2001}
Strathman JG, Kimpel TJ, Dueker KJ, et~al (2001) Bus {Transit} {Operations} {Control}: {Review} and an {Experiment} {Involving} {Tri}-{Met}’s {Automated} {Bus} {Dispatching} {System}. Journal of Public Transportation 4(1):1--26. \doi{10.5038/2375-0901.4.1.1}, \urlprefix\url{https://www.sciencedirect.com/science/article/pii/S1077291X22004398}

\bibitem[{Tang et~al.(2024)Tang, Qu, Jiang, Mo, Cao, Rodriguez, Koutsopoulos, Wu, and Zhao}]{tang_robust_2024}
Tang Y, Qu A, Jiang X, et~al (2024) Robust {Reinforcement} {Learning} {Strategies} with {Evolving} {Curriculum} for {Efficient} {Bus} {Operations} in {Smart} {Cities}. Smart Cities 7(6):3658--3677. \doi{10.3390/smartcities7060141}, \urlprefix\url{https://www.mdpi.com/2624-6511/7/6/141}, number: 6 Publisher: Multidisciplinary Digital Publishing Institute

\bibitem[{Tirachini et~al.(2021)Tirachini, Godachevich, Cats, Muñoz, and Soza-Parra}]{tirachini_headway_2021}
Tirachini A, Godachevich J, Cats O, et~al (2021) Headway variability in public transport: a review of metrics, determinants, effects for quality of service and control strategies. Transport Reviews pp 1--25. \doi{10.1080/01441647.2021.1977415}, \urlprefix\url{https://www.tandfonline.com/doi/full/10.1080/01441647.2021.1977415}

\bibitem[{Wolofsky(2019)}]{wolofsky_towards_2019}
Wolofsky GT (2019) Towards 3-minutes : application of holding and crew interventions to improve service regularity on a high frequency rail transit line. Thesis, Massachusetts Institute of Technology, \urlprefix\url{https://dspace.mit.edu/handle/1721.1/123900}, accepted: 2020-02-28T20:50:05Z Journal Abbreviation: Towards three-minutes : application of holding and crew interventions to improve service regularity on a high frequency rail transit line

\bibitem[{Wood et~al.(2018)Wood, Stasko, Tarte, Jefferson, and Reddy}]{wood_real-time_2018}
Wood D, Stasko T, Tarte L, et~al (2018) A {Real}-{Time} {Service} {Management} {Decision} {Support} {System} for {Train} {Dispatching} at {New} {York} {City} {Transit}. Transportation Research Record 2672(8):327--338. \doi{10.1177/0361198118792116}, \urlprefix\url{https://doi.org/10.1177/0361198118792116}, publisher: SAGE Publications Inc

\end{thebibliography}

\end{document}